\documentclass{osa-article}

\journal{oe}

\usepackage{ulem}

\begin{document}

\title{Spectral-domain optical coherence tomography based on nonlinear interferometers}

\author{Arturo Rojas-Santana,\authormark{1} Gerard J. Machado,\authormark{2}, Maria V. Chekhova,\authormark{3,4}  Dorilian Lopez-Mago,\authormark{1,*} and Juan P. Torres\authormark{2,5,*}}

\address{\authormark{1}Tecnologico de Monterrey, Escuela de Ingenier\'ia y Ciencias, Ave. Eugenio Garza Sada 2501, Monterrey, N.L. 64849, Mexico\\
\authormark{2} ICFO - Institut de Ciencies Fotoniques, The Barcelona Institute of Science and Technology, 08860 Castelldefels (Barcelona), Spain\\
\authormark{3} Max-Planck Institute for the Science of Light, Staudtstr. 2, Erlangen D-91058, Germany\\
\authormark{4} Friedrich-Alexander University of Erlangen-Nuremberg, Staudtstr. 7/B2, Erlangen D-91058, Germany\\
\authormark{5} Departament of Signal Theory and Communications, Universitat Politecnica de Catalunya, 08034 Barcelona, Spain}

\email{\authormark{*}dlopezmago@tec.mx, juanp.torres@icfo.eu} 

\begin{abstract}
Optical coherence tomography (OCT) is a 3D imaging technique that was introduced in 1991 [Science {\bf 254}, 1178 (1991); Applied Optics {\bf 31}, 919 (1992)].  Since 2018 there has been growing
interest in a new type of OCT scheme based on the use of so-called {\it nonlinear interferometers}, interferometers that contain optical parametric amplifiers. Some of these OCT schemes are based on the idea of induced coherence [Physical Review A {\bf 97}, 023824 (2018)], while others make use of an SU(1,1) interferometer [Quantum Science and Technology {\bf 3} 025008 (2018)]. What are the differences and similarities between the output signals measured in standard OCT and in these new OCT schemes? Are there any differences between OCT schemes based on induced coherence and on an SU(1,1) interferometer? Differences can unveil potential advantages of OCT based on nonlinear interferometers when compared with conventional OCT schemes. Similarities might benefit the schemes based on nonlinear interferometers from the wealth of research and technology related to conventional OCT schemes. In all cases we will consider the scheme where the optical sectioning of the sample is obtained by measuring the output signal spectrum ({\it spectral, or Fourier-domain OCT}), since it shows better performance in terms of speed and sensitivity than its counterpart {\it time-domain OCT}.
\end{abstract}

\section{Introduction}
Optical Coherence Tomography (OCT) is a 3D  high-resolution imaging scheme that produces tomographic images of a variety of objects, such as biological systems, by measuring light back-scattered from the samples \cite{drexler2008optical}. To obtain good transverse resolution (in the plane perpendicular
to the beam propagation axis), OCT focuses light into a small spot that is scanned over the sample. To obtain good resolution in the axial
direction (optical sectioning along the beam propagation direction), OCT uses light with a large
bandwidth. OCT is a highly mature optical imaging technology as well as a very active topic of research (for instance, see www.octnews.org for reports in advances in optical coherence tomography).

The first OCT systems were demonstrated in 1991 \cite{Huang1991,dresel1992}, and most of current OCT systems follow the same general structure of these pioneering systems. They use a broadband light beam that splits into two beams in a Michelson interferometric setup: the reference and object beams.  The output signal results from the combination of the reference beam with the object beam after being reflected from the sample. We will refer to these OCT systems as {\it standard} OCT, although there is still a rich variety among these {\it conventional} OCT systems. 

In the last few years several research groups have demonstrated new OCT schemes based on nonlinear interferometers~\cite{Valles2018,paterovaOCT2018,ramelowOCT2020,gerardOCT2020}, interferometers that contain parametric amplifiers~\cite{Chekhova2016}. The main advantage of these quantum schemes is that they allow probing the sample at a chosen wavelength, for instance in the far infrared to achieve higher penetration depth into the sample, while at the same time using an optimum wavelength for efficient detection. Nonlinear interferometers are key elements in numerous applications, namely, in imaging~\cite{Lemos2014,Cardoso2018}, sensing~\cite{Kutas2020}, spectroscopy~\cite{Kalashnikov2016,Paterova2018,Lindner2020}, and microscopy~\cite{Kviatkovsky2020,Paterova2020}. There are two general configurations considered. One scheme in based on the concept of induced coherence~\cite{Mandel1991,Mandel1991_2,Mandel1992}, the other schemes are variants of an SU(1,1) interferometer~\cite{Yurke1986}. We will analyze the OCT signal obtained in both configurations. For the sake of comparison, we will also consider the signal obtained in a {\it standard} OCT scheme.

There are two procedures to obtain the internal axial structure of samples (optical sectioning) in OCT. In {\it time-domain OCT}, each axial scan of the sample consists of the signal measured for an array of different delays introduced in the reference arm of the interferometer. For each delay, the intensity of the signal resulting from the combination of the reference and object beams is measured. The internal structure of the sample is reconstructed from the interferogram obtained plotting the signal measured versus delays.  In {\it spectral, or Fourier-domain OCT} (FD-OCT)~\cite{FERCHER199543} (see also chapter 5 in Ref.~\cite{drexler2008optical}) one retrieves the structure of the sample along the axial direction by Fourier transforming the spectrum of the interferogram with a fixed delay in the reference arm. FD-OCT avoids the need of TD-OCT for displacing mechanically a mirror, and consequently is more robust and allows faster data acquisition. Moreover, it shows better sensitivity than TD-OCT~\cite{choma2003a}. Here we will restrict ourselves to the analysis of OCT signal obtained in FD-OCT in all cases considered. 

One of the main drawbacks of most non-classical implementations of OCT is the low photon flux. This is due to the fact that most implementations work in the low parametric gain regime of optical parametric amplifiers, where the number of photons per mode generated
is much smaller than one \cite{Valles2018,paterovaOCT2018,ramelowOCT2020}. However this low-photon flux can be beneficial in applications that requires a minimal photo dose and where imaging at video rate can still be achieved \cite{gelebart2021}. Higher photon fluxes can be achieved in the high parametric gain regime, where the number of photons per mode is higher than one \cite{gouet2009,gerardOCT2020}. Effects like induced coherence can also be observed in this high parametric regime \cite{belinsky2000,Wiseman2000}. We will consider here a general case, and will derive simplified and useful expressions for the low parametric gain regime of parametric amplifiers.

Finally, one word of caution for the sake of clarity. In 2003 an OCT scheme was demonstrated that was termed as quantum OCT (q-OCT) \cite{Nasr2003}. q-OCT uses paired photons generated in the nonlinear process of spontaneous parametric down-conversion (SPDC), and quantum interference in a Hong-Ou-Mandel (HOM) scheme. Therefore it requires the measurement of  second-order correlation functions that translate into the detection of two-photon coincidences. OCT schemes based on nonlinear interferometers are fundamentally different. While they also make use of paired photons generated in SPDC, they do not require the detection of coincidences and are therefore much simpler.
A comparison of the advantages and disadvantages of both quantum schemes is an interesting topic but it is outside the scope of our present analysis.

\section{Signal measured in standard Fourier-domain optical coherence tomography}
\begin{figure}[t!]
\centering
\includegraphics[width=14 cm]{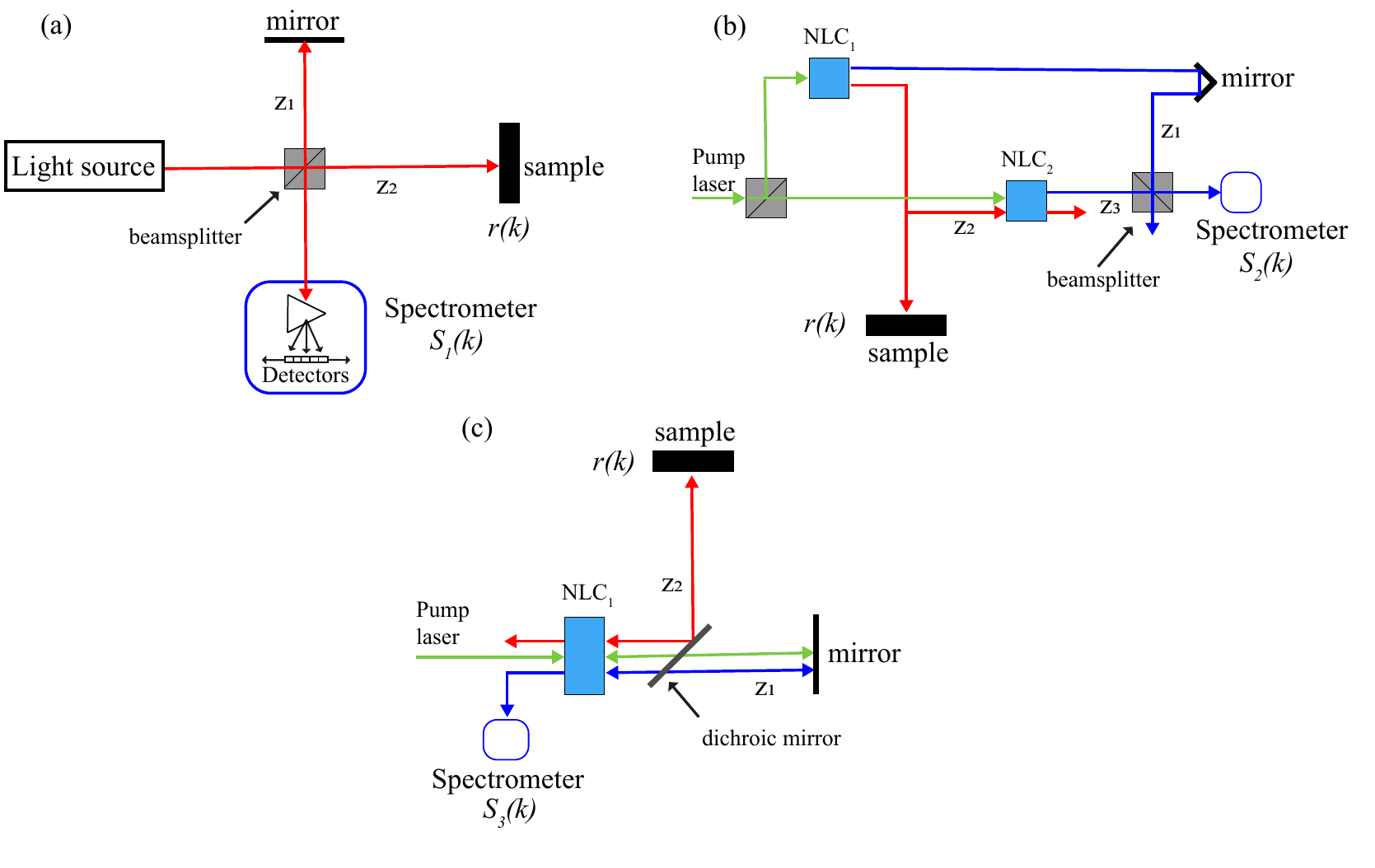}
\caption{Sketch of the three optical coherence tomography (OCT) configurations considered. (a) Standard Fourier-domain (FD-OCT), (b) FD-OCT based on induced coherence, and (c) FD-OCT based on an SU(1,1) interferometer. $S_{i}(k)$ $i=1,2,3$ are the spectral densities measured for each configuration, which are labeled with a different sub-index for the sake of clarity. $r(k)$ is the reflection coefficient of the sample, which depends on its internal layer structure. $z_{1}$, $z_{2}$ and $z_{3}$ are optical paths taken by light beams in each scenario.  Different colors depict light beams (photons) at different central wavelengths. In (b) and (c), $NLC$ refers to nonlinear crystal.}
\label{fig:1}
\end{figure}

For the sake of comparison, we begin by describing a {\it standard} OCT configuration [see Fig. \ref{fig:1}(a)], which will constitute the benchmark for all non-classical OCT schemes (see for instance \cite{drexler2008optical} for an excellent description of OCT and a detailed mathematical derivation of main equations). 

The sample of reference used in the calculations is a single layer with low reflectivity at both faces. Its reflectivity $r(k)$ with respect to the field can thus be written as
\begin{equation} 
\label{sample}
r(k) = r_{1} + r_{2} \exp\left[ 2 i k_0 n_0+2 i k n_g d \right],
\end{equation}
where $R_1=|r_1|^2$ and $R_2=|r_2|^2$ are the reflectivities of the first and second faces,
$n_0$ is the refractive index at the central frequency, $n_g$ is the group index, $d$ is the sample thickness,
$k =\Omega/c$ is the wavenumber deviation from the central wavenumber $k_0 = \omega_0 / c$, and $\Omega$ is the angular frequency deviation from the central frequency $\omega_0$. 

FD-OCT uses light with a large bandwidth: the coherence length, which determines the axial resolution of OCT, is inversely proportional to the spectral bandwidth. We characterize the spectral density of the light source with a Gaussian function
\begin{equation} 
\label{Gaussian-spectrum}
\Phi(k) = \frac{1}{ \pi^{1/2} W} \exp \left( -\frac{k^2}{W^2}\right),
\end{equation}
The function $\Phi(k)$ is normalized so that $\int dk \Phi(k)=1$.  The FWHM spectral bandwidth of the laser source $\Delta \lambda$ is related to the parameter $W$ as $\Delta \lambda=(\lambda_0^2 \sqrt{\ln 2}/\pi) \times  W$, with $\lambda_0$ being the central wavelength of the laser source. For the sake of simplicity, throughout this paper we will consider as main variable the wavenumber, since the axial structure of samples will be obtained from Fourier transforming the output signal written with $k$ as variable.

Light is divided by a beam splitter into the {\it reference} and {\it object} beams in a Michelson interferometer. We consider a $50$:$50$ beam splitter with reflection and transmission coefficients $1/\sqrt{2}$ and $i/\sqrt{2}$, respectively.  The distance traversed by the reference beam is denoted by $2 z_1$, and the distance traversed by the object beam is $2 z_2 = 2(z_1 + s)$. $2s$ is thus the optical path unbalance between the reference and object paths in the Michelson interferometer.  The beams finally recombine at the beam splitter and a  spectrometer measures the spectral density $S_1(k)$ of the output beam, which is given by
\begin{align} \label{S1_exact}
S_1(k) = \frac{\Phi(k)}{4} \bigg| \exp \left[ 2i (k_0 + k) z_1\right] + r(k) \exp \left[2i (k_0 + k) z_2\right]\bigg|^2,
\end{align} 
This is the general expression of the spectral density of the OCT signal for an arbitrary reflection coefficient. After expanding the modulus in Eq. (\ref{S1_exact}), the term independent of $s$ corresponds to {\it self-interference}~\cite{drexler2008optical}, while the two $s$-dependent terms are the {\it cross-interference} terms. For the sample described by Eq. (\ref{sample}), the spectral density is
\begin{eqnarray} 
\label{S1_bilayer}
& & S_1(k)=\frac{1+R_1+R_2}{4}\,\Phi(k) + \frac{2 (R_1 R_2)^{1/2} }{1+R_1+R_2}\, \Phi(k)\,\cos \left[ 2 n_g d\, k +\varphi_1 \right]   \nonumber \\
& & + \frac{2R_1^{1/2} }{1+R_1+R_2}\,\Phi(k)\, \cos \left[ 2 s\, k +\varphi_2  \right] + \frac{2R_2^{1/2}}{1+R_1+R_2}\, \Phi(k)\,\cos \left[ 2s\,k + 2 n_g d\,k +\varphi_3   \right] 
\end{eqnarray}
where $\varphi_1=2 k_0 n_0 d$, $\varphi_2=2 k_0 s$ and $\varphi_3=2k_0 (s + n_0 d)$. 

In order to retrieve information about the sample, e.g., the location of the interfaces and reflection coefficients, we obtain the Fourier transform of the spectral density measured $S_1(k)$, i.e.,  $\hat{S}_1(z)=\mathcal{F}[S_1(k)]= 1/(2\pi)^{1/2}\, \int dk S(k) \exp \left( -ikz\right)$. We find that
\begin{eqnarray} \label{S1hat}
& & \hat{S}_1(z)=\frac{1+R_1+R_2}{4}\, \hat{\Phi}\left(z\right)    \nonumber \\
& &  + \frac{(R_1 R_2)^{1/2} }{4}\, \left[ \hat{\Phi}\left(z - 2n_gd\right)\,\exp\left( i\varphi_1\right)  + \hat{\Phi}\left(z + 2n_gd\right)\, \exp\left(- i\varphi_1\right)  \right] \nonumber \\
& & + \frac{R_1^{1/2}}{4}\, \left[ \hat{\Phi}\left(z - 2 s\right) \,\exp\left(i \varphi_2\right) +  \hat{\Phi}\left(z + 2 s\right)\,\exp\left(-i \varphi_2\right)  \right] \nonumber \\
& & +\frac{R_2^{1/2}}{4}\, \left[ \hat{\Phi}\left(z - 2s-2n_gd)\,\exp\left( i\varphi_3\right) \right)+  \hat{\Phi}\left(z + 2s+2n_g d \right) \,\exp\left( -i\varphi_3\right)\right], 
\end{eqnarray}
where $\hat{\Phi}(z) = \mathcal{F}[\Phi(k)]$. 

In order to identify clearly the location and height of the peaks corresponding to the {\it cross-interference} terms, it is important to have a sufficiently large value of the path-length difference $s$.  If $\Delta z$ is the width of the function $\hat{S}_1(z)$, which determines the axial resolution of OCT, we need $2s \gg \Delta z$, $2s+2n_g d \gg \Delta z$ and $2n_g d \gg \Delta z$. In this case the typical shape of $\hat{S}_1(z)$ contains seven characteristic peaks. A central peak at $z=0$, three peaks located at $z>0$ and another three peaks at $z<0$. The peaks at $z \ne 0$ are located symmetrically around $z=0$. The separation between the two {\it cross-interference} peaks at $z>0$ (or $z>0$) is $2 n_g d$, twice the optical thickness of the sample. For low reflectivity $R_{1,2} \ll 1$, the heights of the {\it cross-interference} in comparison with the height of the peak at $z=0$ are $R_1^{1/2}$ and $R_2^{1/2}$.

\section{Signal measured in Fourier-domain optical coherence tomography in an induced coherence configuration}
Figure \ref{fig:1}(b) shows a sketch of an OCT scheme based on the concept of induced coherence \cite{Valles2018}. Signal ($s_1$) and idler ($i_1$) photons are generated in the first nonlinear crystal (NLC$_1$) by means of spontaneous parametric down-conversion. The signal and idler photons have different central wavelengths, represented in the figure by different colors. The quantum state of the photons generated can be described in the Heisenberg picture by the Bogoliuvov transformations \cite{gatti2003,Dayan2007,torres2011}
\begin{eqnarray}
    & & a_{s_1}(k)=U_{s_1}(k) b_s(k)+V_{s_1}(k) b_i^{\dagger}(-k) \nonumber \\
    & & a_{i_1}(k)=U_{i_1}(k) b_i(k)+V_{i_1}(k) b_s^{\dagger}(-k) \nonumber \\
\end{eqnarray}
where $a_{s_1}$ and $a_{i_1}$ are the operators at the output face of the nonlinear crystal and $b_{s}$ and $b_{i}$ are the operators at the input face. Please see the the appendix for the expressions of the functions $U_{s_1,i_1}$ and $V_{s_1,i_1}$.  

The idler photons are reflected from the sample. The quantum state of the idler photons transforms as
\begin{eqnarray}
    a_{i_1}(k) \Longrightarrow r(k) a_{i_1}(k)+f(k)
\end{eqnarray}
where the operators $f(k)$ fulfill the commutation relationships $\left[ f(k),f^{\dagger}(k^{\prime})\right]=(1-|r(k)|^2) \delta(k-k^{\prime})$. These operators take into account the reflection of the idler photons from a sample with reflectivity $r(k)$ \cite{boyd2008}. After reflection the idler photons are injected into a second nonlinear crystal (NLC$_2$), where signal photons $s_2$, and outgoing idler photons, indistinguisble from the injected idler photons, are generated by means of parametric amplification. Signal photons $s_1$ and $s_2$ are combined on a beam splitter and the spectral density $S_2(k)$ of the resulting interference signal is measured. One can show that the spectral density of the output beam is (see Appendix for details)
\begin{eqnarray} 
\label{S2_exact}
& & S_2(k) =\left| V_{s_2}(k) \right|^2 \bigg[ 1-|r(-k)|^2 \bigg] \nonumber \\
 & & +\bigg|V_{s_1}(k) \exp \left[ i k_s(k) z_1 \right]  + r^{*}(-k) U_{i_1}^*(-k) V_{s_2}(k)  \exp \left[-i k_{i}(-k)z_2+i k_s(k) z_3 \right] \bigg|^2.
\end{eqnarray}
where $z_{1,2,3}$ are distances that signal and idler photons traverse as indicated in Fig. \ref{fig:1}(b). We can write $z_3+z_2=z_1+2s$, so $2s$ is the path unbalance between signal photons $s_1$ and $s_2$ that interfere. This expression is valid in general, for all parametric gain regimes. 

At low parametric gain, we can obtain a simpler expression for the spectrum $S_2(k)$ (see Appendix for further details)
\begin{eqnarray} \label{S2_lowgain}
& & S_2(k)= \left| V_{s_1}(k)\right|^2  \left\{ 1 +R_1^{1/2}  \cos \left[ k \left( 2s+c D_i L \right)- \varphi_1 \right] \right. \nonumber \\
& & \left. + R_2^{1/2}  \cos \left[ k \left( 2s+2 n_g d+c D_i L \right)-\varphi_2 \right] \right\}, 
\end{eqnarray}
where $\varphi_1= k_s^0 z_1+k_i^0 n_i L - k_i^0 z_2 -k_s^0 z_3+\Delta \varphi_p$, $\varphi_2=\varphi_1+2k_i^0\,n_0\, d$ and $\Delta \varphi_p=\varphi_{p_1}-\varphi_{p_2}$ is the phase difference between the two pump beams that pump the two nonlinear crystals. They originate from the same laser, but they can bear different phases. $k_{s,i}^0$ are central wavenumbers at the signal and idler wavelengths, $n_{s,i}$ are the refractive indexes inside the nonlinear crystals, and $D_{s,i}$ are the corresponding inverse group velocities.

The Fourier transform of the spectral density $S_2(k)$ in the low parametric gain regime is
\begin{eqnarray} \label{S2hat}
& & \hat{S}_2(z)=\hat{V}_s(z)  \nonumber\\
& & + \frac{R_1^{1/2}}{2}  \left[ \hat{V}_s\left(z + 2s+cD_iL\right)\, \exp\left( i\varphi_1 \right)+ \hat{V}_s\left(z -2s-cD_iL\right)\, \exp\left( -i\varphi_1 \right)\right] \\
& & + \frac{R_2^{1/2}}{2} \left[ \hat{V}_s\left(z + 2s+2n_gd+cD_iL\right)\,\exp\left(i\varphi_2\right)+ \hat{V}_s\left(z -2s-2n_gd-cD_iL\right)\,\exp\left(-i\varphi_2\right) \right],  \nonumber 
\end{eqnarray}
where $\hat{V}_s(z)$ is the Fourier transform of $\left|V_{s_1}(k)\right|^2$. For observing clear peaks in the output signal $\hat{S}_2$, which allow to determine the position of the interfaces of the sample, it is necessary that $2s+cD_iL \gg \Delta z$ and $2 n_g d \gg \Delta z$.

One important difference with the signal in standard OCT [see Eq. (\ref{S1_exact})] is that the Fourier-transformed spectrum shows only five peaks: one peak at $z=0$, two peaks for $z>0$ and two peaks for $z<0$. The peaks at $z\ne 0$ are located symmetrically around $z=0$. There are no terms equivalent to the {\it self-interference} peaks of standard OCT. The distance between the two peaks at $z>0$ (or $z<0$) is $2n_g d$, which gives the optical thickness of the sample.  

\section{Signal measured in Fourier-domain optical coherence tomography in a SU(1,1) configuration}
Figure \ref{fig:1}(c) shows a sketch of an OCT scheme based on a SU(1,1) interferometer \cite{Paterova2020,ramelowOCT2020,gerardOCT2020}. Signal ($s_1$) and idler ($i_1$) photons generated in the first pass by the nonlinear crystal are separated with the help of a dichroic mirror. Signal photons are reflected and injected back into the same nonlinear crystal. The idler photons are reflected from the sample before travelling back to the nonlinear crystal. After parametric amplification in the second pass of the pump through the nonlinear crystal, the signal photon $s_2$ is sent to a spectrometer. The spectral density of the signal photons is (see Appendix for further details)
\begin{eqnarray}
\label{S3_exact}
S_3(k)&=&\left| V_{s_2}(k) \right|^2 \big[ 1-|r(-k)|^2 \big] \nonumber  \\ 
&+& \bigg|  U_{s_2}(k) V_{s_1}(k)\exp \left[ i \varphi_{s}(k) \right]  + r^{*}(-k) U_{i_1}^*(-k) V_{s_2}(k) \exp \left[-i \varphi_{i}(-k) \right] \bigg|^{2},
\end{eqnarray}
where $\varphi_s(k)=2 (k_s^0+k) z_1$ and $\varphi_i(k)=2 (k_i^0+k) z_2$. $2 z_1$ and $2 z_2$ are the distances traversed by signal and idler photons before entering the nonlinear crystal.

In the low parametric gain regime, the expression of the spectral density simplifies to 
\begin{eqnarray} \label{S3_lowgain}
& & S_3(k) =2\left| V_{s_1}(k)\right|^2  \left\{1 + R_1^{1/2}  \cos \left[ k\left(2s + c D L \right)-\varphi_1 \right] \right. \nonumber \\
& & \left. + R_2^{1/2}  \cos \left[ k\left(2s + 2 n_g d + c D L \right)-\varphi_2\right]\right\},
\end{eqnarray}
where the group velocity mismatch is $D=D_i-D_s$, the path unbalance between signal and idler photons is $2s=2(z_2-z_1)$, $\varphi_1=k_s^0 n_s L+k_i^0 n_i L + 2 k_s^0 z_1+ 2 k_i^0 z_2+\Delta \varphi_p$ and $\varphi_2=\varphi_1 +  2 k_i^0\,n_0\, d$.

The Fourier transform of $S_3(k)$ in the low parametric gain regime is 
\begin{eqnarray} \label{S3hat}
& & \hat{S}_3(z)=2 \hat{V}_{s}(z) \nonumber \\
& & + R_1^{1/2} \left[   \hat{V}_{s}\left(z + 2 s + c D L  \right)\, \exp\left( i\varphi_1\right)+\hat{V}_{s}\left(z - 2 s - c D L  \right)\,\exp\left( -i\varphi_1 \right) \right]  \\
& & + R_2^{1/2} \left[ \hat{V}_{s}\left(z + 2 s + 2 n_g d + c D L \right)\,\exp\left( i\varphi_2\right) +  \hat{V}_{s}\left(z - 2 s - 2 n_g d - c D L \right)\,\exp\left( -i\varphi_2\right) \right], \nonumber 
\end{eqnarray}
where $\hat{V}_{s}(z)$ is the Fourier transform of $\left|V_{s_1}(k)\right|^2$. Again the Fourier-transformed spectrum shows five peaks. The signal for OCT in an SU(1,1) scheme is very similar to the OCT signal for an induced coherence scheme. The main difference is that now, for observing clear peaks in the output signal $\hat{S}_3$, it is necessary that $2s+cDL \gg \Delta z$ where $D=D_i-D_s$, in contrast to the case of induced coherence that is   $2s+cD_iL \gg \Delta z$. Since $D_i \ll D$, in an induced coherence scheme the requirement of large path unbalance is not so stringent.  

\section{Discussion}
\begin{figure}[t!]
\centering
\includegraphics[width=12.0cm]{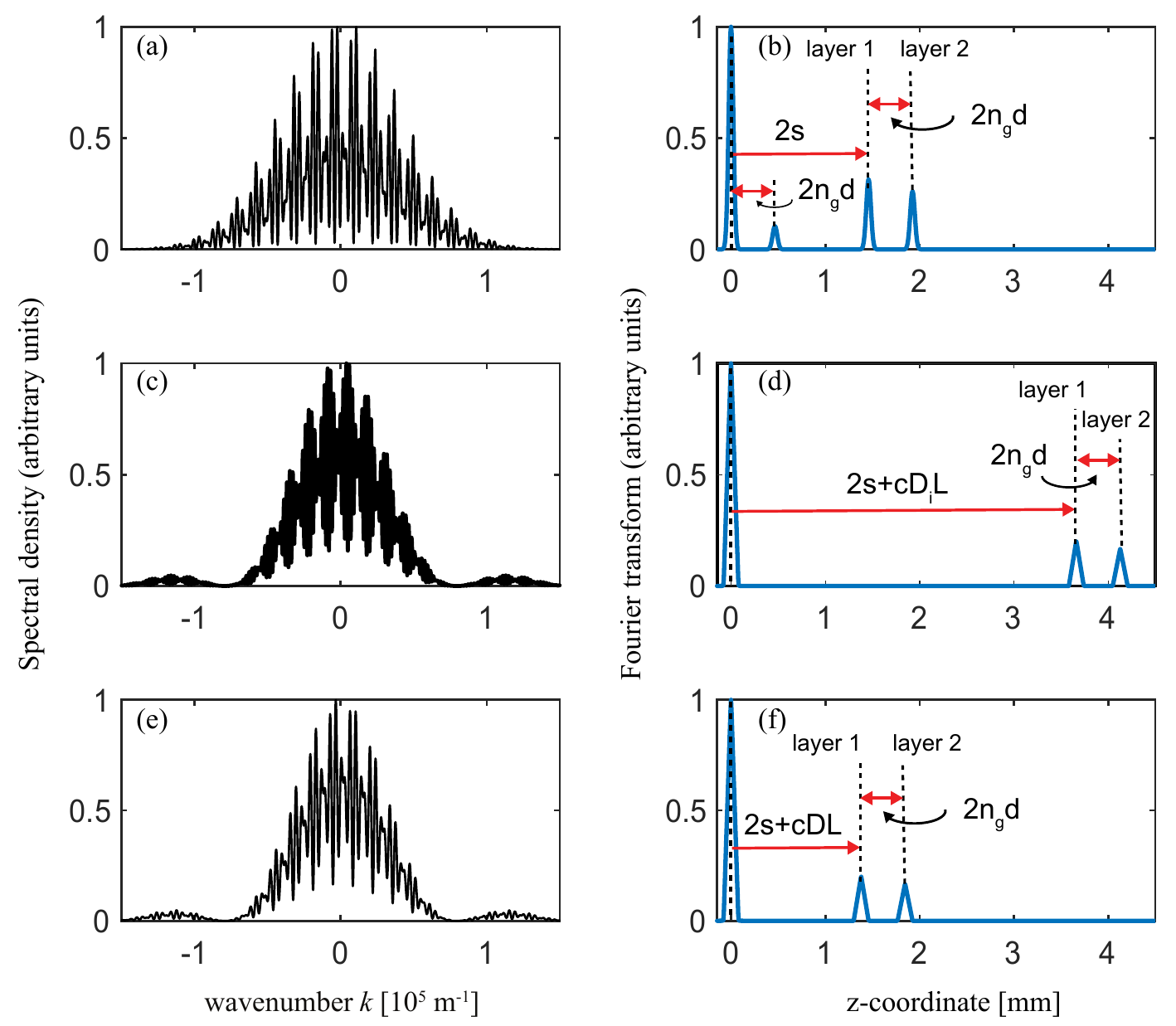}
\caption{The spectral density and its Fourier transform in the three OCT schemes considered. (a), (c) and (e): Spectral density $S_i(k)$ for $i=1,2,3$, and (b), (d) and (f): Modulus of the Fourier transform of the spectral density $\hat{S}_i(z)$.
(a) and (b) correspond to standard OCT;  (c) and (d) correspond to OCT based on an induced coherence scheme in the low parametric gain regime; and (e) and (f) consider the case of OCT based on a SU(1,1) interferometer in the low parametric gain regime. The sample thickness of the bilayer structure considered is $d=100\mu$m, the refractive index ($n_0$) and group index ($n_g$) are equal to $2.33$ and the path unbalance is $s=730\mu$m. The spectral density and its Fourier transform are depicted in arbitrary units.}
\label{fig:2}
\end{figure}

Figure \ref{fig:2} shows the spectrum $S_i(k)$ ($i=1,2,3$) and its Fourier transform $|\hat{S}_i(z)|$ for the three cases considered here in the low parametric gain regime. Due to the symmetry of the figure, we plot only the signal for $z\ge 0$. In standard OCT we consider a light beam with central wavelength $\lambda = 810n$m and bandwidth $\Delta \lambda=10n$m. For the quantum schemes, we consider a MgO-doped Lithium Niobate crystal with length $L=1$mm and parameters $cDL = -79.1\mu$m and $cD_iL=2.2$mm. The wavelength of the signal and idler photons generated in PDC are $\lambda_s^0=810$ nm and $\lambda_i^0=1550$ nm. The wavelength of the pump beam is $\lambda_p=532$ nm. We consider a layer with thickness $d=100\mu$m, refractive index ($n_0$) and group refractive index ($n_g$) equal to $2.33$. The optical thickness is thus $n_0 d=0.233\mu$m. We consider reflection coefficients $R_1=0.16$ and $R_2=0.11$ for the first and second interfaces of the sample.

The main difference between standard OCT and the two other schemes is that the signal obtained in standard OCT shows peaks corresponding to {\it self-interference} and {\it cross-reference terms}, while there are no {\it self-interference} terms in the two other cases. When comparing the OCT signals based on induced coherence and an SU(1,1) interferometer, we observe that the only appreciable difference between them is that the peaks at $z \ne 0$ are further away from the central peak at $z=0$ for the case of induced coherence. The reason for this is that the peaks are located at $2s+cDL$ for the $SU(1,1)$ configuration, and at $2s+cD_i L$ for the induced coherence configuration. 

Eqs. (\ref{S1_exact}), (\ref{S2_exact}) and (\ref{S3_exact}) are exact expressions that give the spectral density that should be obtained in the three OCT systems considered, for a sample with an arbitrary reflection coefficient $r(k)$. To retrieve the sought-after internal structure of the sample under investigation, one needs to Fourier transform these expressions. Since most experiments take place in the low parametric gain regime, we have also derived simplified expressions (\ref{S2_lowgain}) and (\ref{S3_lowgain}) for the spectral density that should be obtained in this regime. We have done it for the specific benchmark sample we are considering. 

The question arises how the shape of the OCT signal expected in the high parametric regime, given by Eq. (\ref{S3_exact}) in general, compares with the equivalent signal in the low parametric gain regime, given by Eq. (\ref{S3hat}). We consider the case of OCT based on a SU(1,1) scheme, where OCT experiments has been reported in both the low (\cite{paterovaOCT2018,ramelowOCT2020}) and high \cite{gerardOCT2020} parametric gain regimes. 

\begin{figure}[t!]
\centering
\includegraphics[width=12.0cm]{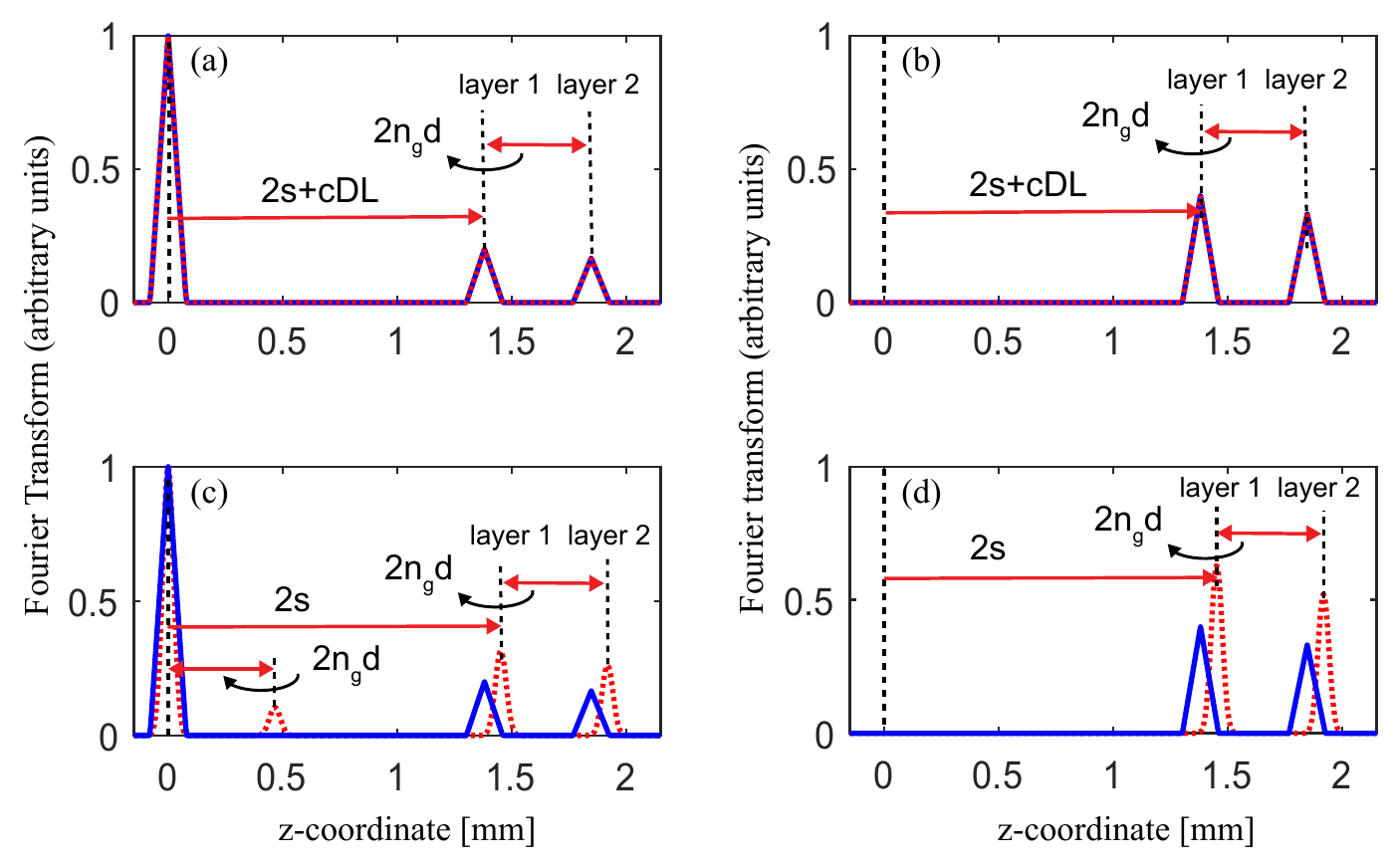}
\caption{Fourier transform of the spectral density for OCT based in a SU(1,1) configuration for two values of the parametric gain. Solid blue lines show the
analytical approximation valid for the low parametric gain regime given by Eq. (\ref{S3hat}) for a value of the gain $G = 0.01$. Dotted red lines correspond to the
exact expression obtained by Fourier transforming Eq. (\ref{S3_exact}).  (a) and (b):
G=0.01 (low parametric gain); (c) and (d) G=10. We consider a nonlinear crystal with parameters $D_s = 7.61 \times 10^{-9} $s/m and $D_i = 7.34 \times 10^{-9}$s/m. We have $cDL = -79.1\mu$m and $cD_iL=2.2$mm. The reflection coefficients are $R_1=0.16$ and $R_2=0.11$ for the first and second interfaces of the single-layer sample, respectively. Panels (b,d) show the result of the subtraction of two measurements with the phases in the reference arm differing by $\pi$.}
\label{fig4_helena}
\end{figure}

Figs. \ref{fig4_helena} (a) and (c) show the Fourier transform of the expression given by Eq. (\ref{S3_exact}) (dotted red lines) and the analytical approximation of the Fourier transform (solid blue lines) given by Eq. (\ref{S3hat}) and valid in the low parametric gain regime. As expected both expressions give the same result for a low gain of $G=0.01$ [Fig. \ref{fig4_helena} (a)]. For large gain [$G=10$, Fig. \ref{fig4_helena} (c)], we still observe the two peaks corresponding to the two interfaces of the single layer separated a distance $2n_g d$. However the amplitudes of the peaks of the Fourier transform change and do not have a straightforward relationship with the reflectivity of the layers of the sample as it is the case in the low parametric gain regime. Moreover a new peak at a distance $2 n_g d$ appears that can make difficult, in principle, to resolve the axial structure of the sample. All parametric regimes share the presence of an intense DC component that otherwise bears no relevant information about the sample.

In standard OCT it has been demonstrated \cite{wojtkowski2002,choma2003b,leitgeb2003b} that one can obtain a signal without the self-interference and DC terms by subtracting the spectral densities obtained in two measurements. In one of these measurements, a $\pi$ phase is introduced into the beam propagating in the reference arm. We apply this procedure to the signals $S_3(k)$ shown in Figs. \ref{fig4_helena} (a) and (c). In our case we need to subtract two spectral densities given by  Eq. (\ref{S3_exact}) with phases $\varphi_s(k)$ and $\varphi_s(k)+\pi$, and from Eq. (\ref{S3_lowgain}) with phases $\varphi_1(k)$ and $\varphi_1(k)+\pi$. Figs. \ref{fig4_helena} (b) and (d) show the results. The central peak is removed as expected. But also the new peak that appears for $G=10$ is removed, so in this respect this peak behaves similarly to an self-interference term in standard OCT.  

\begin{figure}[t!]
\centering
\includegraphics[width=8.0cm]{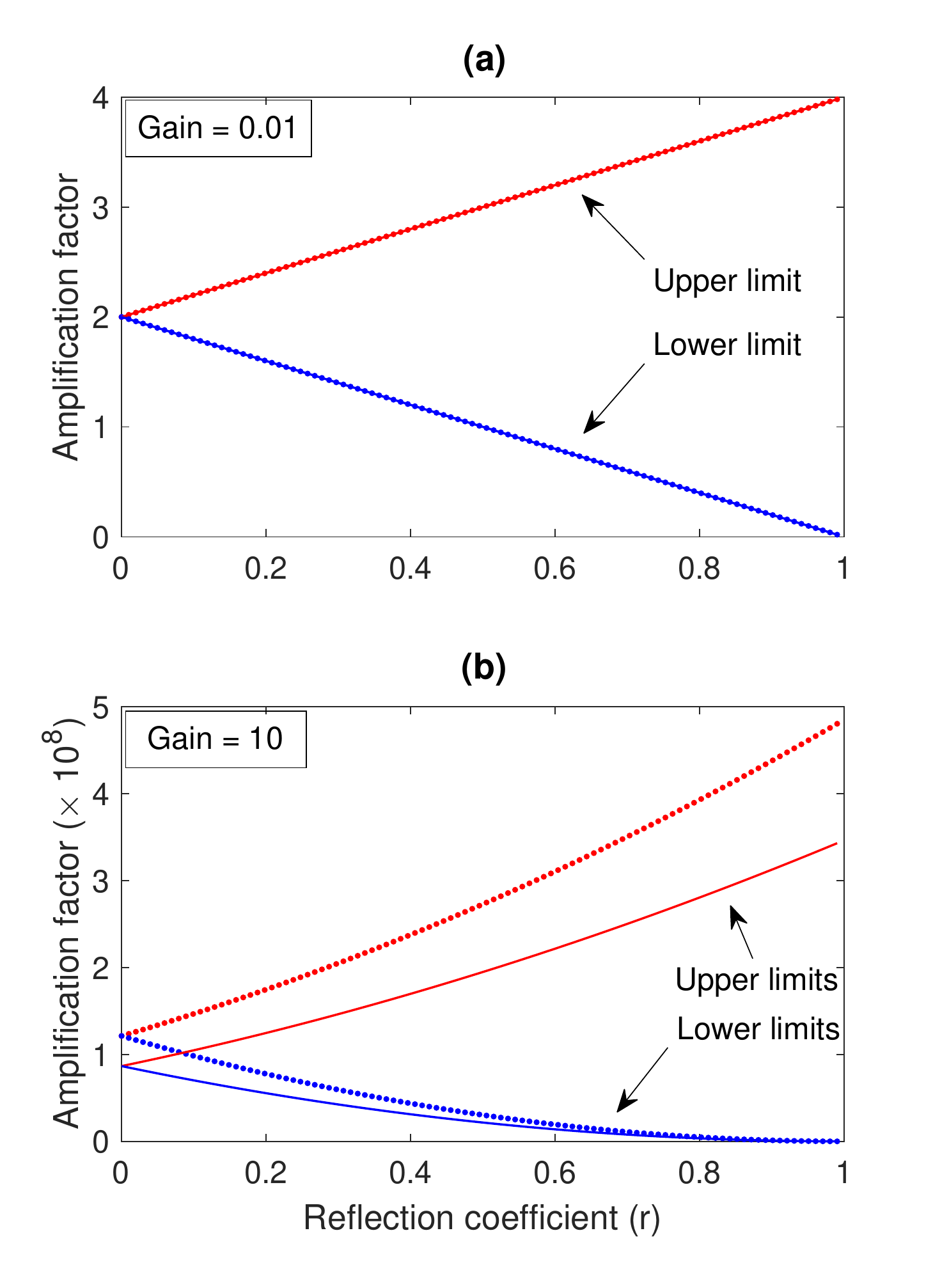}
\caption{The amplification factor $\gamma$, i.e., the ratio of the signal power at the detection stage and the idler power probing the sample as a function of the single-interface reflectivity $r$  for two values of the parametric gain $G$. (a) $G=0.01$ (low parametric gain) and (b) $G=10$ (high parametric gain). Dashed lines: exact solution. Solid lines: Results obtained using the single-mode approximation. Red lines: Maximum value of the amplification factor for a given $r$. Blue lines: minimum value of the amplification factor for a given $r$.}
\label{fig2_helena}
\end{figure}

In certain applications, for instance in biological imaging and in art conservation studies, the power of the light beam that interacts with the sample is low to avoid damage to the sample if too many photons are absorbed. An important advantage of the SU(1,1) configuration in the high parametric gain regime, when compared with standard and induced coherence OCT schemes, is that the flux rate of photons that interacts with the sample, $N_{i_1}(t)=\langle a_{i_1}^{\dagger}(t) a_{i_1}(t) \rangle$ is much lower than the flux rate of photons that are detected, $N_{s_2}(t)=\langle a_{s_2}^{\dagger}(t) a_{s_2}(t) \rangle$. We define the amplification factor as $\gamma=N_{s_2}(t)/N_{i_1}(t)$. Although we consider explicitly time $t$, the flux rate of photons is constant since we assume that the pump beam is CW light. We should notice that even though the theoretical analysis in the high parametric gain regime consider a CW pump with the corresponding nonlinear coefficient, in experiments one needs to use a pulsed pump with high energy pulses that are highly focused. Notwithstanding the approximation of CW and plane wave pump beam is still valid \cite{Dayan2007,spasibko2012}. In Fig. \ref{fig2_helena} we plot the maximum (red dotted line) and minimum (blue dotted line) values of the amplification factor $\gamma$ as a function of the reflectivity of a single-layer sample $r$. 

To get further insight, we consider the single-mode description of OCT based on an SU(1,1) interferometer. This is equivalent to considering a single frequency $k=0$ in the spectral density given by Eq. (\ref{S3_exact}). We obtain that the flux rate of photons interacting with the sample is 
\begin{equation}
N_{i_1}=|V|^2
\end{equation} 
and the flux rate of signal photons detected is
\begin{equation}
    N_{s_2}=|V|^2 \left[ (1-|r|^2)+ (1+|r|^2)|U|^2+2|r||U|^2 \cos \theta \right].
\end{equation}
Here $|U|=|U_{s}(k=0)|=|U_{i}(k=0)|$, $|V|=|V_{s}(k=0)|=|V_{i}(k=0)|$ and $\theta$ summarizes the phases that appear in Eq. (\ref{S3_exact}). The maximum and minimum values of the amplification factor in the single-mode approximation are:
\begin{eqnarray}
& & \text{Max}(\gamma)=1-|r|^2+(1+|r|)^2 |U|^2 \nonumber \\ & & \text{Min}(\gamma)=1-|r|^2+(1-|r|)^2 |U|^2.
\end{eqnarray}
These values are also plotted in Fig. \ref{fig2_helena} as solid lines (red depicts the maximum and blue the minimum value).

For low parametric gain, the approximate (single-mode) and exact values of the amplification factor are barely distinguishable. For high parametric gain we observe that the {\it simple} consideration of single-mode parametric amplification provides a very good approximation to the exact values of the amplification factor. For low reflectivity the amplification factor can be approximated as $\gamma=1+|U|^2=1+\cosh^2 G$ and does not depend on the phase $\theta$. For high reflectivity the amplification factor can vary between $0$ and $\gamma=4|U|^2=4\cosh^2 G$.

For comparison, in the case of induced coherence  the flux rate of photons interacting with the sample is, similarly to the case of an SU(1,1) interferometer, $N_{i_1}=|V|^2$. Now there are two signal beams: the flux rate of signal photons generated in the first nonlinear crystal is $s_1$ is $N_{s_1}=|V|^2$, while the flux rate of signal photons generated in the second nonlinear crystal is $N_{s_2}=|V|^2(1+|r|^2 |V|^2)$.

\section{Conclusion}
We have analyzed the signal of interest, which enables the optical sectioning of samples in axial scans, in three different OCT schemes. For the sake of comparison, we considered {\it standard} OCT and two other schemes that make use of nonlinear interferometers. One is based on the concept of induced coherence, another one on an SU (1,1) interferometer.  We have considered Fourier-domain OCT, where the axial internal structure of the sample is obtained from the Fourier transform of the signal. Our analysis provides an overview of tomographic images acquired with new OCT schemes that make use of nonlinear interferometers.

\begin{figure}[t!]
\centering
\includegraphics[width=10.0cm]{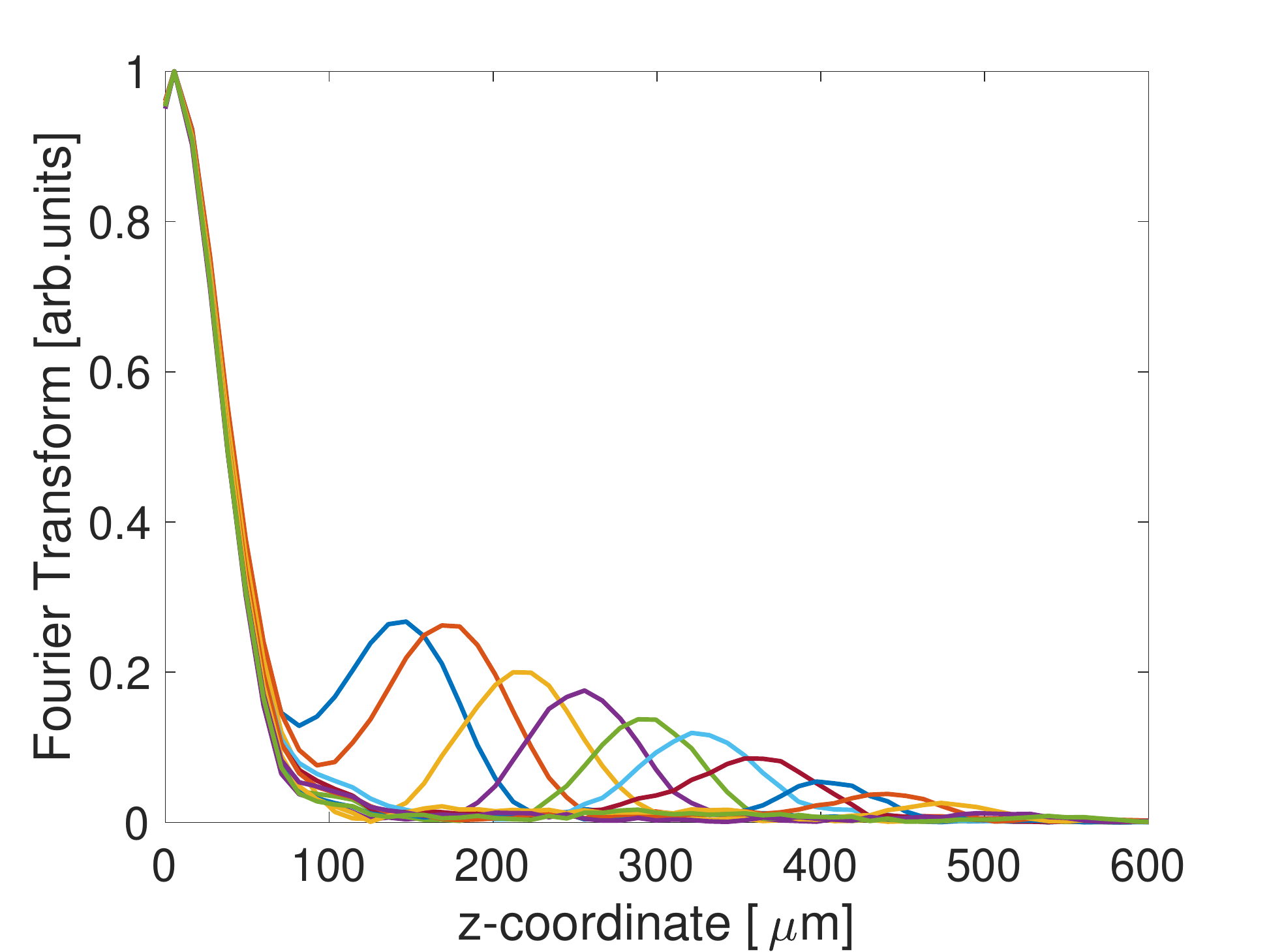}
\caption{{\bf Sensitivity decay in an OCT scheme based on a SU(1,1) interferometer}. The idler photons are reflected from a mirror with ideal reflectivity $r=1$ located at different axial positions. The spectral density of signal photons is measured for each position, and we plot here the Fourier transform of the spectral density, which shows the location of the mirror and its reflectivity. The heights of all peaks would be the same for ideal {\it point-like} pixels. However, the finite size of the pixels at the detection stage causes an apparent decay of the value of the reflectivity. Experimental data were obtained from \cite{gerardOCT2020}. Different colors correspond to different positions of the mirror.}
\label{fig_gerard.eps}
\end{figure}

The signal in {\it standard} OCT shows both unwanted {\it self-interference} terms and {\it cross-interference} terms, which carry the sought-after information about the sample. A distinguishing characteristic of OCT schemes based on nonlinear interferometers in the low parametric gain regime is that they do not show any  {\it self-interference} terms. However, there are techniques that make use of the phases that accompany each term to remove {\it self-reference} terms, and even the peak at $z=0$ \cite{wojtkowski2002,leitgeb2003b,choma2003b}. When considering these techniques, which have been successfully demonstrated in several OCT schemes, the signal obtained in the three OCT schemes turns out to be essentially the same for the representative sample consider here. We expect a similar behaviour for other types of samples.

The only difference remaining is in the location of peaks of interest at $z\ne 0$. For our benchmark example, in the low parametric gain regime, the peaks at $z>0$ corresponding to the first interface, after Fourier transform,  are located at $2s$, $2s+c D_i L$ and $2s+cDL$, for the three OCT schemes considered respectively. This is clearly seen in Fig. \ref{fig:2}. This puts different conditions on the path unbalance $s$ required to obtain a clear image of the internal structure of the sample in Fourier-domain OCT.

OCT based on nonlinear interferometers has the advantage that the sample can be probed at one wavelength, while detection occurs at another wavelength. Both wavelengths can be chosen independently to optimize probing and detection. 

Finally, all OCT schemes might share similar technological limitations. For instance, it is well known that the finite size of pixels in spectrometers in general, and in high-sensitivity CCD cameras in particular, leads to a sensitivity decay as a function of path unbalance \cite{leitgeb2003a,yun2003,nassif2004,zhilin2007}. This sensitivity decay needs to be corrected for obtaining high-resolution and accurate images of samples. Fig. \ref{fig_gerard.eps} shows an example of this sensitivity decay for an OCT scheme based on an SU(1,1) interferometer with gain $G=1.7$. This experimental data is obtained from \cite{gerardOCT2020}. This is an example that shows that the wealth of research associated to {\it standard OCT} schemes can be also beneficial for new OCT schemes based on nonlinear interferometers.

\section{Appendix: The quantum state of paired photons generated in spontaneous parametric down-conversion}
\subsection{The quantum state of signal-idler paired photons}
Here we derive the expression for the quantum state of paired photons generated in spontaneous parametric down-conversion using the Heisenberg picture.

A nonlinear crystal (length $L$) is pumped by an intense CW laser beam with wavelength $\lambda_p$. The wavenumber of the pump beam is $k_p=2\pi n_p/\lambda_p$ with $n_p$ being the refractive index at $\lambda_p$. The interaction of the pump beam with the crystal mediates the generation of pairs of correlated photons (signal and idler) by means of spontaneous parametric down-conversion. The paired photons generated are frequency anti-correlated, i.e. $k_{s}=k_{s}^0+k$ and $k_{i}=k_{i}^0-k$, where $k_{s,i}$  are the wavenumbers of signal/idler photons, $k_{s,i}^0$ are the corresponding central wavenumbers, and $\pm k$ is the wavenumber deviation from the corresponding central wavenumber. 

We assume that the bandwidth of the photons is much larger than the bandwidth of the pump beam. We also assume that the Rayleigh length ($L_R=k_p^0 w_p^2/2$) of the pump beam is much larger than the crystal length. Here $w_p$ is the pump beam waist and $k_p^0=2\pi n_p /\lambda_p^0$ is the pump beam wavenumber.  Under these conditions we can describe the spatio-temporal characteristics of parametric down-conversion using the CW and plane pump beam approximation. The Bogoliuvov transformations that relate the quantum operators for signal and idler photons ($a_s$ and $a_i$) at the output face of the nonlinear crystal to the quantum operators at the input face of the nonlinear crystal ($b_s$ and $b_i$) are \cite{gatti2003,Dayan2007,torres2011}
\begin{eqnarray}
& & a_s(k)=U_s(k) \,b_{s}(k) + V_s(k) \,b_{i}^{\dagger}(-k), \\
& & a_i(k)=U_i(k)\, b_{i}(k) + V_i(k)\,  b_{s}^{\dagger}(-k), 
\end{eqnarray}
where
\begin{eqnarray}
& & U_{s,i}(k)=  \left\{ \cosh (\Gamma_{s,i} L)-i\frac{\Delta_{s,i}}{2\Gamma_{s,i}} \sinh (\Gamma_{s,i} L) \right\} \exp \left\{ i\left[ k_p+k_{s,i}(k)-k_{i,s} (-k) \right] \frac{L}{2}\right\}, \label{Eq:Usi}  \\
& & V_{s,i}(k)= -i \frac{\sigma}{\Gamma_{s,i}}  \sinh (\Gamma_{s,i} L) \exp \left\{ i\left[ k_p+k_{s,i}(k)-k_{i,s} (-k) \right] \frac{L}{2}+i \varphi_p\right\}. \label{Eq:Vs}
\end{eqnarray}
$\varphi_p$ is the phase of the pump beam, the nonlinear coefficient $\sigma$ is
\begin{equation}
\sigma=\left( \frac{\hbar \omega_p \omega_s \omega_i [\chi^{(2)}]^2 F_0}{8\epsilon_0 c^3 S n_p n_s n_i} \right)^{1/2},
\end{equation}
and
\begin{equation}
\Gamma_{s,i}=\left[ \sigma^2-\frac{\Delta_{s,i}^2}{4} \right]^{1/2}
\end{equation}
$F_0=P_0/(\hbar \omega_p)$ is the flux rate density of pump photons (photons/s), $S$ is the area of the pump beam and $P_0$ is the pump power. The low parametric gain regime corresponds to a gain $G=\sigma L \ll 1$.

The signal and idler beams fulfill the paraxial approximation, so we can expand the corresponding wavenumbers in a Taylor series as $k_{s,i}=k_{s,i}^0+cD_{s,i} k$. $D_{s,i}$ are inverse group velocities. We assume phase matching at the central frequencies, i.e., $k_p=k_s^0+k_i^0 \pm 2\pi/\Lambda$, where $\Lambda$ is the poling period of the nonlinear crystal.  The phase matching function is thus $\Delta_s=-\Delta_i=cDL k$ with $D=D_i-D_s$. 

In the low parametric gain approximation,
\begin{eqnarray}
& & U_{s,i}(k)=\exp \left[ i k_{s,i}(k) L \right] \nonumber \\
& & V_{s,i}(k)=-i \,(\sigma L)\, \text{sinc} \frac{\Delta_s L}{2}  
\exp \left\{ i \left[ k_p+k_{s,i}(k)-k_{i,s}(-k)\right] \frac{L}{2}+i \varphi_p\right\}.
\label{PDC}
\end{eqnarray}

\subsection{Description of the effect on the quantum state of idler photons of reflection from a sample with frequency-dependent reflectivity}
The idler beam interacts with a sample with reflectivity $r(k)$. The operator that describes the quantum state of idler photons after reflection from the sample is \cite{boyd2008}
\begin{equation}
a_i(k) \Longrightarrow r(k) a_i(k)+f(k),
\end{equation}
where $f(k)$ fulfills the commutation relationship 
\begin{equation}
[f(k),f^{\dagger}(k^{\prime})]= \left[1- |t(k)|^2\right]\,\delta(k-k^{\prime}).
\label{commutation}
\end{equation}
The operator $f(k)$ takes into account the frequency-dependent losses and reflectivity of the sample.

\section*{Funding}
We acknowledge support from the Spanish Ministry of
Economy and Competitiveness (Severo Ochoa program
for Centres of Excellence in R\&D No. SEV-2015-0522),
from Fundacio Privada Cellex, from Fundacio Mir-Puig,
and from Generalitat de Catalunya through the CERCA
program. This work was also funded through the
EMPIR project 17FUN01-BeCOMe. The EMPIR initiative is co-funded by the European Union Horizon 2020
research and innovation programme and the EMPIR participating States. D.L.M. acknowledges funding from Consejo Nacional de Ciencia y Tecnolog\'{i}a (Grants No. 293471, No. 295239, and No. APN2016-3140). 

\section*{Acknowledgments}
A.R.S. acknowledges support from Becas de Movilidad CONACYT. GJM was supported by the Secretaria d’Universitats i Recerca del Departament d’Empresa i Coneixement de la Generalitat de Catalunya and European Social Fund (FEDER).

\section*{Disclosures}
The authors declare no conflicts of interest.

\bibliography{JorgeArturo2021}

\end{document}